\documentclass[aps,prd,preprint,superscriptaddress,amsmath,amssymb,showpacs]{revtex4-1}
\usepackage{dcolumn}
\usepackage{graphicx}
\usepackage{float}
\usepackage{physics}
\usepackage[colorlinks=true,allcolors=blue]{hyperref}

\begin{document}

\title{Energy loss of heavy and light quarks in holographic magnetized background}

\author{Zhou-Run Zhu}
\affiliation{College of Science, China Three Gorges University, Yichang 443002, China}

\author{Sheng-Qin Feng}
\email{fengsq@ctgu.edu.cn}
\affiliation{College of Science, China Three Gorges University, Yichang 443002, China}
\affiliation{Key Laboratory of Quark and Lepton Physics (MOE) and Institute of Particle Physics,\\
Central China Normal University, Wuhan 430079, China}

\author{Ya-Fei Shi}
\affiliation{College of Science, China Three Gorges University, Yichang 443002, China}

\author{Yang Zhong}
\affiliation{College of Science, China Three Gorges University, Yichang 443002, China}

\date{\today}

\begin{abstract}
We systematically study holographic effects on the magnetic field dependence of the drag force, diffusion coefficient, jet quenching parameter of heavy quarks, and the shooting string energy loss of light quarks in the RHIC and LHC energy regions by using the AdS/CFT correspondence in this paper. This study is motivated by the phenomena of a strong magnetic field and jet quenching, which have been found in relativistic heavy ion collisions. The probe's direction of motion is perpendicular and parallel to the direction of magnetic field $B$. The effects of a magnetic field on energy loss when moving perpendicular to the magnetic field direction are larger than moving parallel to the magnetic field direction, which implies that the magnetic field tends to suppress more quarks and jets when moving in the transverse direction than in the parallel direction. It is found that the diffusion coefficient decreases with the magnetic field in the transverse direction, but increases with the magnetic field in the parallel direction, which indicates that the quark may diffuse farther when moving parallel to the magnetic field direction. We also find that the magnetic field will enhance the energy loss of the light quarks when moving in the transverse direction than in the parallel direction.
\end{abstract}


\maketitle

\section{Introduction}\label{sec:01_intro}
 It is well known that the experiments at the Relativistic Heavy Ion Collider (RHIC) and the Large Hadron Collider (LHC) can generate a new state of matter so-called quark gluon plasma (QGP)\cite{Adams:2005dq,Adcox:2004mh,Shuryak:2004cy,Aamodt:2010pa}. One of the important properties of QGP is jet quenching: when a high energy parton propagates through the medium, the presence of the medium in which the energetic parton finds itself has two significant effects: it causes the parton to lose energy and it changes the direction of the parton’s momentum \cite{Wang:1991xy}. In the study of relativistic heavy ion collisions, besides jet quenching, another important phenomenon is the generation of a strong magnetic field of noncentral heavy ion collisions at the RHIC and the LHC\cite{Voronyuk:2011jd,Skokov:2009qp,Bzdak:2011yy,Deng:2012pc,Mo:2013qya,She:2017icp,Zhong:2014cda}. Besides the phenomenological significance in theoretical sides of strong interaction, strong magnetic fields in relativistic heavy-ion collisions also provide some deep investigations of the dynamics of quantum chromodynamics (QCD), of which the vacuum structure is of numerous interests.

 In order to study the important features of jet quenching，which defines the energy loss as a high-energy parton passes through the QGP during RHIC and LHC energy regions, one should make it work by passing some colored objects through the finite temperature $N = 4$ super-Yang-Mills plasma. Readiest to hand is an external quark, which in the framework of AdS/CFT\cite{Witten:1998qj,Gubser:1998bc,Maldacena:1997re} is represented as a string dangling from the boundary of $\mathrm{AdS}_5$-Schwarzschild black hole\cite{Gubser:2006bz,Rey:1998ik,Brandhuber:1998bs,Maldacena:1998im,Rey:1998bq}. Its point of attachment on the boundary carries a fundamental charge under the gauge group $SU(N)$, and it is infinitely massive. The magnetic field is introduced into the background geometry by solving the Einstein-Maxwell system\cite{Li:2016gfn,Dudal:2015wfn,Chelabi:2015gpc}. After embedding the magnetized background geometry into the modified soft-wall model, the magnetic field dependent behaviors of jet quenching and energy loss can be worked out numerically.

 AdS/CFT correspondence or the gauge/string duality has yielded many important explorations into the dynamics of strongly coupled gauge theories\cite{Brustein:2008cg,Demir:2008tr,Pourhassan:2016icn,Herzog:2006gh,Sadeghi:2010zp}.  The jet quenching parameter for $N = 4$ SYM plasma has been carried out by Liu $\textit{et al.}$ in their seminal work\cite{Liu:2006ug}.  On the other hand, gauge/string duality can be taken as an insightful tool in this issue and many quantities have already been studied, such as heavy quarkonium entropy\cite{Chen:2017lsf}, shear viscosity to entropy density ratio\cite{Critelli:2014kra} and energy loss\cite{Mamo:2016xco,Zhang:2018mqt,Finazzo:2016mhm,Rougemont:2015wca}. A new study for light quark energy loss by using shooting strings in SYM plasma to overcome the difficulties of previous falling string holographic scenarios was carried out by the authors of Refs.\cite{Ficnar:2013qxa,Ficnar:2013wba,Chesler:2008wd,Abelev:2012pa}.

The study of the QCD features for strong magnetic field $B$ has attracted considerable attention\cite{Evans:2010xs,Alam:2012fw,Preis:2010cq,Ballon-Bayona:2013cta,Bali:2011qj,Kharzeev:2012ph}. Recently, the jet quenching parameter in a strongly coupled $N = 4$ SYM plasma with a strong magnetic field was analyzed in\cite{Li:2016bbh,Zhang:2018pyr}. The inverse magnetic catalysis was analyzed in\cite{Mamo:2015dea}. The main motivation for studying the magnetic field holographic QCD features is that strong magnetic field $\mathfrak{B}$($\mathfrak{B}(0.02-0.25) \textrm{GeV}^{2}$) is generated in relativistic heavy ion collisions experiments at the RHIC and the LHC\cite{Voronyuk:2011jd,Skokov:2009qp,Bzdak:2011yy,Deng:2012pc,Mo:2013qya,She:2017icp,Zhong:2014cda}, which has amounted in effects on the QGP created during these heavy ion collision experiments\cite{Kharzeev:2007jp,Abelev:2009ac,Selyuzhenkov:2011xq}.

 Holographic effects on the magnetic field dependence of the drag force, diffusion coefficient, jet quenching parameter of heavy quarks and the shooting string energy loss of light quarks are studied in the RHIC and LHC energy regions by using the AdS/CFT correspondence in this article. The paper is organized as follows. In Sec.~\ref{sec:02}, we introduce the gravity background with a backreaction of a magnetic field through the Einstein-Maxwell (EM) system. In Secs.~\ref{sec:03} and ~\ref{sec:04}, we calculate the drag force and diffusion coefficient. The jet quenching parameter is computed in Sec.~\ref{sec:05}. The shooting string energy loss of the light quarks is computed in Sec.~\ref{sec:06}. A short discussion and conclusion are given in Sec.~\ref{sec:07}.

\section{Background geometry}\label{sec:02}
The gravity background with a backreaction of a magnetic field through the Einstein-Maxwell system was introduced by\cite{Li:2016gfn,Dudal:2015wfn,Mamo:2015dea}, and the action is as follows:
\begin{equation}
\label{eq1}
\ S = \frac{1}{16\pi G_5 }\int \dd{x}^5{ \sqrt{- g }(L-F^{MN}F_{MN}+\frac{12}{R^{2} })},
\end{equation}
where $L$ represents the scalar curvature, $G_{5}$ represents $5D$ Newton constant, $g$ is the determinant of metric $g_{\mu\nu}$, $R$ represents the AdS radius, and $F_{MN}$ is the U(1) gauge field \cite{Li:2016gfn}.

Since we assume that the magnetic field is along the $x_{3}$ axis, which breaks the rotation symmetry, the ansatz for metric can be taken as
\begin{equation}
\label{eq2}
\ ds^{2}=\frac{R^{2}}{z^{2}}[-f(z)dt^{2}+h(z)(dx_{1}^{2}+dx_{2}^{2})+q(z)dx_{3}^{2}+\frac{dz^{2}}{f(z)}].
\end{equation}

When $z=z_{h}$, a black hole solution is $f(z=z_{h})=0$.

It has been pointed out \cite{Mamo:2015dea} that the perturbative solution can work well when $B\ll T^{2}$. The physical magnetic field $\mathfrak{B}$ at the boundary is related with $B$ that we use in this article by the equation $\mathfrak{B}=\sqrt{3}B$. We take $B\leq 0.15 \textrm{GeV}^{2}$, and the physical magnetic field $\mathfrak{B}\leq 0.26 \textrm{GeV}^{2}$ which conforms to the RHIC and LHC magnetic field. As a first approximation, one can only take the leading expansion in\cite{Li:2016gfn} as
\begin{equation}
\label{eq10}
  f(z) = 1- \frac{z^{4}}{z_{h}^{4}} \left( 1+\frac{2}{3} B^{2} z_{h}^{4}\log(\frac{z}{z_{h}}) \right ),
 \end{equation}
 \begin{equation}
 \label{eq11}
  q(z) = 1+ \frac{2}{3} B^{2}\ln(\mu z)z^{4},
 \end{equation}
 \begin{equation}
 \label{eq12}
  h(z) = 1- \frac{1}{3} B^{2}\ln(\mu z)z^{4},
\end{equation}
where $z_{h}$ is the horizon of a black hole, and $\mu$ is a constant and varies from 0.25-1 GeV, we take it as 0.5 GeV in this article. Since the element $q(z)$ is not equal to $h(z)$, it allows us to study the anisotropic cases in the presence of the magnetic field.  The Hawking temperature with magnetic field $B$ is calculated as
\begin{equation}
\label{eq13}
T=\frac{1}{\pi z_{h}}-\frac{B^{2}z_{h}^{3}}{6\pi}.
\end{equation}

The Hawking temperature $T=T(z_{h})$, which is a function of the position of the horizon, corresponds to the temperature of the thermal bath in the gauge theory. In this article, we will expand this method to study the effects on the magnetic field and temperature of the drag force, diffusion coefficient, jet quenching parameter of heavy quarks, and the shooting string energy loss of light quarks.

\section{Drag force in a magnetized background}\label{sec:03}
 When a heavy quark moves though the strongly coupled QGP with an unchanging velocity $v$, it loses energy via drag force from the trailing string model \cite{Gubser:2006bz,Herzog:2006gh}. It can also be analyzed by the Langevin coefficient \cite{Giataganas:2013hwa,Giataganas:2013zaa}. When a quark moves on the boundary of the AdS, the end of the string hangs down into the bulk of the AdS space. The shape of the trailing string does not change with time when the quark has been moving at a constant velocity for a long time. The drag force is correlated with the damping rate $\mu$ (or friction coefficient), and the Langevin equation is
\begin{equation}
\label{eq14}
 \frac{dp}{dt}=-\mu p+f_{1},
\end{equation}
where $f_{1}$ is the driving force. The driving force is equal to the drag force for $dp/dt=0$.
The effects on the magnetic field of the drag force while moving perpendicular and parallel to the magnetic field are discussed as follows.

\subsection{Moving perpendicular to the magnetic field}
When the heavy quark is moving perpendicularly to the magnetic field, the coordinates are parametrized by
\begin{equation}
\label{eq15}
\ t=\tau,\ x_{1}=vt+\xi(z),\  x_{2}=x_{3}=0,\ z=\sigma.
\end{equation}

A classical trailing string can be described by the Nambu-Goto action
\begin{equation}
\label{eq16}
 S= - \frac{1}{2\pi\alpha'}\int d \sigma d\tau  \sqrt{- \det g_{\alpha\beta}},
\end{equation}
where $g_{\alpha\beta}$ is the determinant of the induced metric,
\begin{equation}
\label{eq17}
\ g_{\alpha\beta} = g_{\mu\nu}\frac{\partial X^{\mu}}{\partial \sigma^{\alpha}}\frac{\partial X^{\nu}}{\partial \sigma^{\beta}},
\end{equation}
where $g_{\mu\nu}$ and $X^{\mu}$ are the brane metric and target space coordinates, respectively.

By plugging Eq.(\ref{eq15}) into Eq.(\ref{eq2}), one gets the metric as
\begin{equation}
\label{eq18}
 \ d{s}^2 = g_{tt}\ d{t}^2 + g_{xx}\dd{x_1}^2 + g_{zz}\dd{z}^2,
\end{equation}
and the induced metric is
\begin{equation}
\label{eq19}
\ g_{tt}=-\frac{R^{2}}{z^{2}}f(z),\ g_{xx}=\frac{R^{2}}{z^{2}}h(z),\ g_{zz}=\frac{R^{2}}{z^{2}}\frac{1}{f(z)}.
\end{equation}

The Lagrangian density is given by
\begin{equation}
\begin{split}
\label{eq20}
\mathcal{L}&= \sqrt{-g_{zz}g_{tt}-g_{zz}g_{xx}\upsilon^{2}-g_{tt}g_{xx}\xi'^{2}}\\
       &=\sqrt{(\frac{R^{2}}{z^{2}})^{2}-(\frac{R^{2}}{z^{2}})^{2}\frac{h(z)}{f(z)}\upsilon^{2}
       +(\frac{R^{2}}{z^{2}})^{2}h(z)f(z)\xi'^{2}},
\end{split}
\end{equation}
where $\xi'= d\xi/d\sigma$. From Eq.(\ref{eq20}), one finds that the action does not rely on $\xi$ explicitly, which means $\partial \mathcal{L}/\partial\xi'$ is a constant. The canonical momentum is given as
\begin{equation}
\label{eq21}
\ \Pi_{\xi}=\frac{\partial \mathcal{L}}{\partial\xi'}=-\xi'\frac{g_{tt}g_{xx}}{\sqrt{-g_{\alpha\beta}}}=
\frac{\xi'^{2}(\frac{R^{2}}{z^{2}})^{2}h(z)f(z)}{\sqrt{(\frac{R^{2}}{z^{2}})^{2}-(\frac{R^{2}}{z^{2}})^{2}\frac{h(z)}{f(z)}\upsilon^{2}
       +(\frac{R^{2}}{z^{2}})^{2}h(z)f(z)\xi'^{2}}},
\end{equation}
which leads to
\begin{equation}
\label{eq22}
\xi'^{2}=\frac{\Pi_{\xi}^{2}[1-\frac{h(z)}{f(z)}\upsilon^{2}]}{h(z)f(z)
[(\frac{R^{2}}{z^{2}})^{2}h(z)f(z)-\Pi_{\xi}^{2}]}.
\end{equation}

It is emphasized that ${\xi'}^{2}$ must be positive everywhere, therefore, the numerator and denominator on the right-hand side of Eq.(\ref{eq22}) are both positive for small $z$ and negative for large $z$ near the horizon. With these, the critical point $z_{c}$ can be written as
\begin{equation}
\label{eq23}
\ h(z_{c})\upsilon^{2}=f(z_{c}),
\end{equation}
and
\begin{equation}
\label{eq24}
\Pi_{\xi}^{2}=(\frac{R^{2}}{z_{c}^{2}})^{2}f(z_{c}) h(z_{c}).
\end{equation}

The current density for momentum $p_{1}$ along the $x_{1}$ direction can be given as
\begin{equation}
\label{eq25}
\pi^{r}_{x_{1}}=-\frac{1}{2\pi\alpha'}\xi'\frac{g_{tt}g_{xx}}{-g_{\alpha\beta}}.
\end{equation}

The the drag force can be obtained as
\begin{equation}
\label{eq26}
f_{\perp}=\frac{dp_{1}}{dt}=\sqrt{-g_{\alpha\beta}}\pi^{r}_{x_{1}}=-\frac{1}{2\pi\alpha'} \Pi_{\xi}=-\frac{1}{2\pi\alpha'}\frac{R^{2}}{z_{c}^{2}}h(z_{c})\upsilon,
\end{equation}
where the minus sign denotes the opposite direction of momentum.

The drag force of the isotropic SYM result at zero magnetic field \cite{Gubser:2006bz} was given by
\begin{equation}
\label{eq27}
f_{SYM}= -\frac{\pi T^{2} \sqrt{\lambda}}{2}  \frac{v}{\sqrt{1-v^{2}}},
\end{equation}
where $ \sqrt{\lambda}=\sqrt{g^{2}_{YM}N_{c}}=\frac{R^{2}}{\alpha'}$.

\subsection{Moving parallel to the magnetic field}
When the heavy quark moves parallel to the magnetic field, the corresponding coordinates are parametrized by
\begin{equation}
\label{eq28}
\ t=\tau,\ x_{3}=vt+\xi(z),\  x_{1}=x_{2}=0,\ z=\sigma.
\end{equation}

By repeating the previous deduction, one can get the drag force as
\begin{equation}
\label{eq29}
f_{\parallel}=-\frac{1}{2\pi\alpha'} \frac{R^{2}}{z_{c}^{2}}q(z_{c})\upsilon.
\end{equation}

\subsection{The calculated results of drag force}
In order to study the anisotropic effect on the magnetic field dependence of the drag force, we study the anisotropic drag force $f/f_{\textrm{SYM}}$  in the magnetized  background normalized by the isotropic SYM result at zero magnetic field in Fig. 1. It is found that the drag force is generally stronger when the heavy quark is moving perpendicular to the magnetic field direction than when it is moving parallel to the magnetic field direction, which indicates that the magnetic field tends to suppress more quarks in the transverse than in the parallel direction.
\begin{figure}
    \centering
    \includegraphics[width=8.5cm]{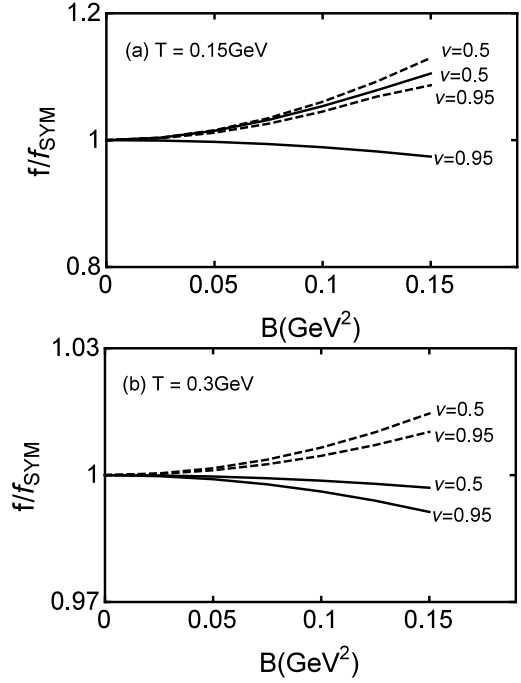}
    \caption{\label{fig1} Anisotropic effect on the magnetic field dependence of the drag force normalized by the isotropic SYM result at zero magnetic field (a) for a low temperature ($T = 0.15 \textrm{GeV}$) and (b) for a higher one ($T = 0.3 \textrm{GeV}$). The solid line indicates the heavy quark moving parallel to the magnetic field direction, and the dashed line denotes the heavy quark moving perpendicular to the magnetic field direction.}
\end{figure}

\begin{figure}
    \centering
    \includegraphics[width=8.5cm]{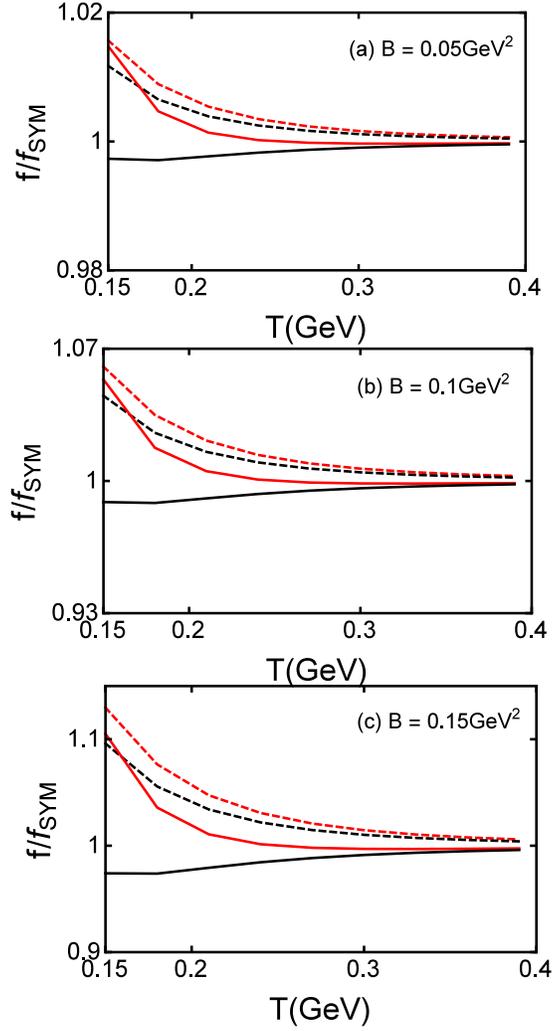}
    \caption{\label{fig2} Anisotropic effect on temperature dependence of the drag force in the magnetized background normalized by the isotropic SYM result at zero magnetic field (a) for $B = 0.05 \textrm{GeV}^2$, (b) for $B = 0.1 \textrm{GeV}^{2}$, and (c) for $B = 0.15 \textrm{GeV}^{2}$. The solid line indicates the heavy quark moving parallel to the magnetic field direction, and the dashed line indicates it moving perpendicular to the magnetic field direction. The red line denotes  a low velocity ($\nu = 0.5$) and the black line denotes a higher one ($\nu = 0.95$).}
\end{figure}

The anisotropic effect on temperature dependence of the drag force $f/f_{\textrm{SYM}}$ in the magnetized background is given in Fig.~\ref{fig2}. It is found that the drag force in the magnetized background is obviously temperature dependent. When the temperature is relatively low ($T \leq 0.3 \textrm{GeV}$), the magnetic field can change the isotropic phase while both moving parallel and transverse to the magnetic field, but at relatively high temperature($T > 0.3$), $f/f_\textrm{SYM}$ tends to be 1, which suggests the magnetic field hardly changes the isotropic phase. The anisotropy $f/f_\textrm{SYM}<1$  is computed in the case of moving parallel to the magnetic field with ultrafast motion($v = 0.95$).

\section{Diffusion coefficient in a magnetic field background}\label{sec:04}
In this section, we focus on the study of the diffusion coefficient which can be derived from the drag force.

The drag force with zero magnetic field in the Eq.(\ref{eq27}) can be rewritten as
\begin{equation}
\label{eq30}
f_{SYM}= -\frac{\pi T^{2} \sqrt{\lambda}}{2m}  \frac{vm}{\sqrt{1-v^{2}}}= -\eta_{D}p,
\end{equation}
where $m$ represents the mass of the heavy quark, $\eta_{D}$ is the drag coefficient, and $p=\frac{mv}{\sqrt{1-v^{2}}}$ is the momentum.

The diffusion time $t_\textrm{SYM}$ is given as
\begin{equation}
\label{eq31}
t_{SYM}= \frac{1}{\eta_{D}}= \frac{2m}{\pi T^{2} \sqrt{\lambda}},
\end{equation}
and the diffusion coefficient can be given by
\begin{equation}
\label{eq32}
D_{SYM}=\frac{T}{m} t_{SYM}=\frac{2}{\pi T \sqrt{\lambda}}.
\end{equation}

When the quark is moving perpendicular to the magnetic field, Eq.(\ref{eq26}) can be rewritten as
\begin{equation}
\label{eq33}
f_{\perp}=-\frac{h(z_{c}) \sqrt{1-v^{2}}}{\pi^{2} T^{2}z_{c}^{2} } \frac{\pi T^{2} \sqrt{\lambda}}{2m}  \frac{vm}{\sqrt{1-v^{2}}},
\end{equation}
and the diffusion time $t_{\perp}$  is given as
\begin{equation}
\label{eq34}
t_{\perp}= \frac{2m}{\pi T^{2} \sqrt{\lambda}} \frac{\pi^{2} T^{2}z_{c}^{2}}{h(z_{c}) \sqrt{1-v^{2}}}.
\end{equation}

The diffusion coefficient can be given by
\begin{equation}
\label{eq35}
D_{\perp}=\frac{T}{m} t_{\perp}=\frac{2}{\pi T \sqrt{\lambda}} \frac{\pi^{2} T^{2}z_{c}^{2}}{h(z_{c}) \sqrt{1-v^{2}}}.
\end{equation}

From Eqs.(\ref{eq32}) and (\ref{eq35}), one can get
\begin{equation}
\label{eq36}
\frac{D_{\perp}}{D_\textrm{SYM}}= \frac{\pi^{2} T^{2}z_{c}^{2}}{h(z_{c}) \sqrt{1-v^{2}}}.
\end{equation}

Then the diffusion coefficient moving parallel to the magnetic field direction can be given as
\begin{equation}
\label{eq37}
D_{\parallel}=\frac{2}{\pi T \sqrt{\lambda}} \frac{\pi^{2} T^{2}z_{c}^{2}}{q(z_{c}) \sqrt{1-v^{2}}}.
\end{equation}

\begin{figure}
    \centering
    \includegraphics[width=8.5cm]{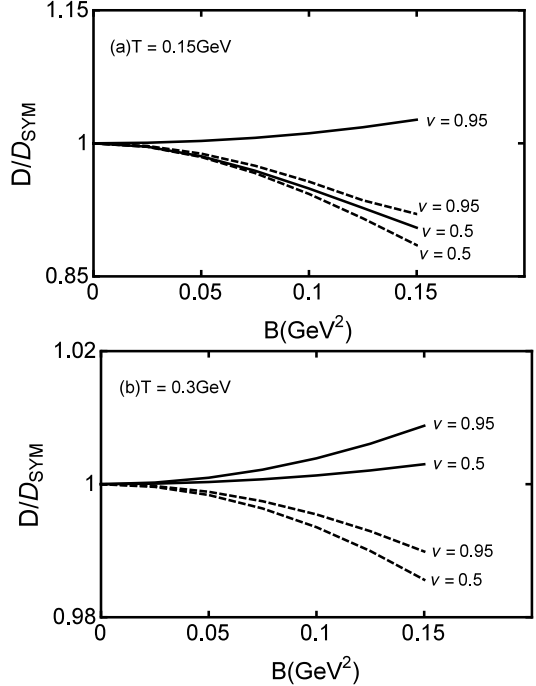}
    \caption{\label{fig3} Anisotropic effect on the magnetic field dependence of the diffusion coefficient normalized by the isotropic SYM result at zero magnetic field (a) for a low temperature ($T = 0.15 \textrm{GeV}$) and (b) for a higher one ($T = 0.3 \textrm{GeV}$). The solid line indicates the heavy quark moving parallel to the magnetic field direction, and the dashed line denotes it moving transverse to the magnetic field direction.}
\end{figure}

The anisotropic diffusion coefficient $D/D_\textrm{SYM}$ in the magnetized background is given in Fig.~\ref{fig3}. It is found that the diffusion coefficient decreases with the magnetic field when moving perpendicular to the magnetic field direction, but increases with the magnetic field when the probe is moving parallel to the magnetic field direction in the high temperature and high velocity, which indicates that the quark may diffusion farther when moving parallel to the magnetic field direction.

\begin{figure}
    \centering
    \includegraphics[width=8.5cm]{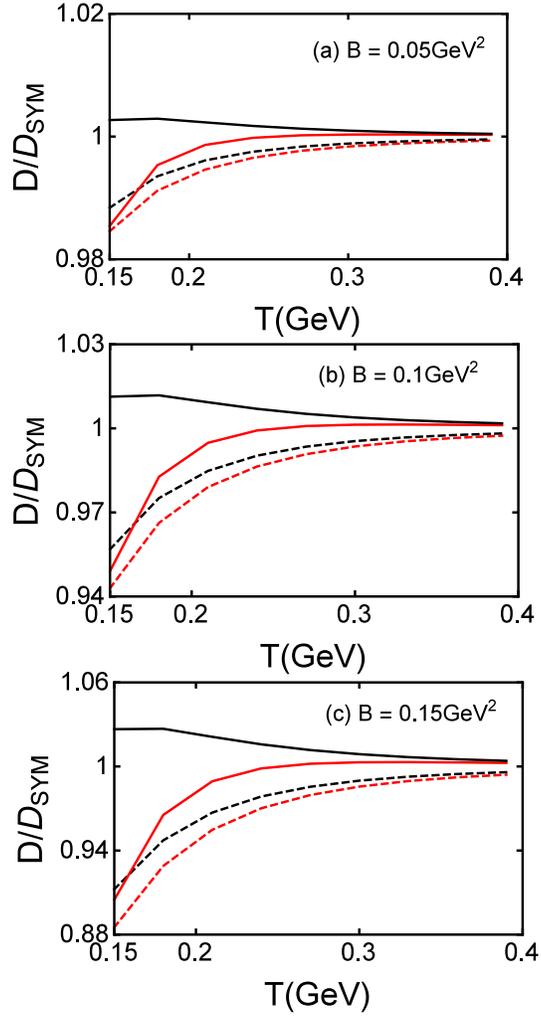}
    \caption{\label{fig4} Anisotropic effect on temperature dependence of the diffusion coefficient in the magnetized background normalized by the isotropic SYM result at zero magnetic field (a)  for $B = 0.05 \textrm{GeV}^2$, (b) for $B = 0.1 \textrm{GeV}^{2}$, and (c) for $B = 0.15 \textrm{GeV}^{2}$. The solid line indicates the heavy quark moving parallel to the magnetic field direction, and the dashed line denotes it moving perpendicular to the magnetic field direction. The red line denotes a low velocity ($\nu = 0.5$) and the black line denotes a higher one ($\nu = 0.95$).}
\end{figure}

The anisotropic effect on the temperature dependence of the diffusion coefficient $D/D_\textrm{SYM}$ in the magnetized  background is given in Fig.~\ref{fig4}. One can see that the diffusion coefficient in the magnetized background is obviously temperature dependent. When the temperature is relatively low ($T \leq 0.3 \textrm{GeV}$), the magnetic field will change the isotropic phase, but at relatively high temperature ($T > 0.3$), $D/D_\textrm{SYM}$ tends to be 1 with both moving parallel and transverse to the magnetic field, which means the magnetic field hardly changes the isotropic phase at relatively high temperature. A special result with the anisotropy of $D/D_\textrm{SYM}>1$ is shown in the case of moving parallel to the magnetic field with ultrafast motion ($v = 0.95$).

\section{Jet quenching parameter in magnetic field background}\label{sec:05}
The strong jet quenching and the elliptic flow\cite{Adler:2003kt,Ackermann:2000tr,Kolb:2001qz} suggest that the matter of relativistic heavy ion collisions is strongly interacting.  AdS/CFT correspondence has produced many important insights into the dynamics of strongly coupled gauge theories.  Liu, Rajagopal and Wiedemann \cite{Liu:2006ug} have studied the jet quenching parameter $\hat{q}$ for $N = 4$ SYM plasma.  Motivated by this, there are many attempts to study the jet quenching parameter in this approach. For example, the effects of an electromagnetic field and the chemical potential on  $\hat{q}$ have been analyzed in\cite{Lin:2006au,Armesto:2006zv,Avramis:2006ip,Sadeghi:2010df,BitaghsirFadafan:2010zh}. The anisotropy effects on $\hat{q}$ are investigated in\cite{Giataganas:2012zy}, and some works\cite{Sadeghi:2012xb,Zhang:2012jd,Fadafan:2008uv} have used some of the results and formulas of \cite{Giataganas:2012zy} to apply them on other anisotropic theories etc.

In this section we will discuss the jet quenching parameter $\hat{q}$ in the magnetized background. It is well known that the jet quenching parameter $\hat{q}$ is related to the Wilson loop with the following equation\cite{Liu:2006ug}:
\begin{equation}
\label{eq38}
\langle W^{A}[C] \approx exp(-\frac{1}{4\sqrt{2}}\hat{q}L^{-}L^{2}),
\end{equation}
where $W^{A}[C]$ is the adjoint Wilson loop and $C$ is a rectangular contour of size $L \times L^{-}$. The quark and antiquark are separated by a small $L$ and travel along the $L^{-}$.

Meanwhile, one can use the following equations
\begin{equation}
\label{eq39}
\langle W^{A}[C]\rangle \approx \langle W^{F}[C]\rangle^{2},
\end{equation}
and
\begin{equation}
\label{eq40}
\langle W^{F}[C]\rangle \approx exp[-S_{I}],
\end{equation}
where $W^{F}[C]$ is the fundamental representation of the Wilson loop and $S_{I} = S - S_{0}$ ($S$ is the total energy of the quark and antiquark pair, and $S_{0}$ is the self-energy of the isolated quark and antiquark).

Then the general relation of the jet quenching parameter is given as
\begin{equation}
\label{eq41}
\hat{q}=8\sqrt{2}\frac{S_{I}}{L^{-}L^{2}}.
\end{equation}

\subsection{Moving perpendicular to the magnetic field}
There are two cases for the probe moving perpendicular to magnetic field direction, the first case is that the jet moves along the $x_{1}$ direction and the momentum broadening occurs along the $x_{3}$ direction(also the magnetic field direction), and the second is that the jet moves along the $x_{1}$ direction and the momentum broadening occurs along the $x_{2}$ direction.

By using the the light-cone coordinate $x^{\mu} = (z, x^{+}, x^{-}, x_{2}, x_{3})$, we start with the first case. One can rewrite metric(\ref{eq2}) as
\begin{equation}
\label{eq42}
\begin{split}
ds^{2}=&\frac{R^{2}}{z^{2}} \bigg\{-\frac{f(z)}{2}[(dx^{+})^{2}+(dx^{-})^{2}]+\frac{h(z)}{2}[(dx^{+})^{2}+(dx^{-})^{2}]
-f(z)dx^{+}dx^{-}\\
&-h(z)dx^{+}dx^{-}+h(z)dx^{2}_{2}+q(z)dx^{2}_{3}\bigg\}+\frac{R^{2}}{z^{2} f(z)}dz^{2},
\end{split}
\end{equation}
then chooses the following static gauge coordinate
\begin{equation}
\label{eq43}
\ x^{-}=\tau,\ x_{3}=\sigma,\  x^{+}=x_{2}=const.
\end{equation}

If one adopts a profile of $z = z(\sigma)$, then Eq. (\ref{eq42}) becomes
\begin{equation}
\label{eq44}
\ ds^{2}=\frac{R^{2}}{2z^{2}}[h(z)-f(z)]d\tau^{2}+\frac{R^{2}}{z^{2}}[q(z)+\frac{\dot{z}}{f(z)}d\sigma^{2}],
\end{equation}
where $\dot{z}=\frac{dz}{d\sigma}$.

The Nambu-Goto action can be given as
\begin{equation}
\label{eq45}
\ S=\frac{\sqrt{2}L^{-}}{2\pi\alpha'} \int^{\frac{L}{2}}_{0}d\sigma\frac{R^{2}}{z^{2}}\sqrt{[h(z)-f(z)][q(z)+\frac{\dot{z}}{{f(z)}}]},
\end{equation}
and the Lagrangian density can be written as
\begin{equation}
\label{eq46}
\ \mathcal{L}=\frac{R^{2}}{z^{2}}\sqrt{[h(z)-f(z)][q(z)+\frac{\dot{z}}{{f(z)}}]}.
\end{equation}

Since the Lagrangian density does not rely on $\sigma$ explicitly, one has a conserved quantity as
\begin{equation}
\label{eq47}
\ \frac{\partial \mathcal{L}}{\dot{z}} - \mathcal{L}=E,
\end{equation}
where $E$ is the constant energy of motion. One can derive the following formulas
\begin{equation}
\label{eq48}
\ \dot{z}=\sqrt{\frac{f(z)q(z)}{E^{2}}} \sqrt{(\frac{R^{2}}{z^{2}})^{2}q(z)[h(z)-f(z)]-E^{2}}.
\end{equation}

The separation parameter $L$ of a quark and antiquark pair can be given by
\begin{equation}
\label{eq49}
\begin{split}
\ L&=2\int^{z_{h}}_{0}dz\cdot \frac{1}{\dot{z}}\\
  &=2\int^{z_{h}}_{0}dz\frac{E}{\sqrt{q(z)f(z)}}\frac{1}{\sqrt{(\frac{R^{2}}{z^{2}})^{2}q(z)[h(z)-f(z)]}}
  [1-\frac{E^{2}}{(\frac{R^{2}}{z^{2}})^{2}q(z)[h(z)-f(z)]}]^{-\frac{1}{2}}.
\end{split}
\end{equation}

In the low limit of $E$, Eq.(\ref{eq49}) turns into
\begin{equation}
\label{eq50}
\ L=2\int^{z_{h}}_{0}dz\frac{E}{\sqrt{q(z)f(z)}}\frac{1}{\sqrt{(\frac{R^{2}}{z^{2}})^{2}q(z)[h(z)-f(z)]}}
  [1+\frac{E^{2}}{2(\frac{R^{2}}{z^{2}})^{2}q(z)[h(z)-f(z)]}].
\end{equation}

By ignoring $E^{2}$, one gets
\begin{equation}
\begin{split}
\label{eq51}
\ L&=2E\int^{z_{h}}_{0}dz\frac{1}{q(z)}\frac{z^{2}}{R^{2}\sqrt{f(z)[h(z)-f(z)]}}\\
   &=2EI,
\end{split}
\end{equation}
where
\begin{equation}
\label{eq52}
\ I=\int^{z_{h}}_{0}dz\frac{1}{q(z)}\frac{z^{2}}{R^{2}\sqrt{f(z)[h(z)-f(z)]}}.
\end{equation}

Substituting Eq.(\ref{eq48}) into Eq.(\ref{eq45}) in the small limit $E$, one calculates the Nambu-Goto action as
\begin{equation}
\label{eq53}
\ S=\frac{\sqrt{2}L^{-}}{2\pi\alpha'} \int^{z_{h}}_{0}dz\sqrt{\frac{1}{f(z)}[h(z)-f(z)](\frac{R^{2}}{z^{2}})^{2}}
[1+\frac{E^{2}}{2(\frac{R^{2}}{z^{2}})^{2}q(z)[h(z)-f(z)]}].
\end{equation}

The self-energy $S_{0}$ is
\begin{equation}
\label{eq54}
\ S_{0}=\frac{\sqrt{2}L^{-}}{2\pi\alpha'} \int^{z_{h}}_{0}dz\sqrt{\frac{1}{f(z)}[h(z)-f(z)](\frac{R^{2}}{z^{2}})^{2}}.
\end{equation}

By subtracting Eq.(\ref{eq54}) from Eq.(\ref{eq53}), one can get
\begin{equation}
\begin{split}
\label{eq55}
\ S&=S-S_{0}\\
    &=\frac{\sqrt{2}L^{-}}{2\pi\alpha'} \int^{z_{h}}_{0}dz\sqrt{\frac{1}{f(z)}[h(z)-f(z)](\frac{R^{2}}{z^{2}})^{2}}
    \frac{E^{2}}{2(\frac{R^{2}}{z^{2}})^{2}q(z)[h(z)-f(z)]}\\
    &=\frac{\sqrt{2}L^{-}E^{2}}{4\pi\alpha'}\int^{z_{h}}_{0}dz\frac{1}{q(z)}\frac{z^{2}}{R^{2}\sqrt{f(z)[h(z)-f(z)]}}\\
   &=\frac{\sqrt{2}L^{-}E^{2}}{4\pi\alpha'}I.
\end{split}
\end{equation}

By inserting Eqs.(\ref{eq51}) and (\ref{eq55}) into Eq.(\ref{eq41}), one can get the jet quenching parameter as
\begin{equation}
\begin{split}
\label{eq56}
\hat{q}_{\perp(\parallel)}&=8\sqrt{2}\frac{S_{I}}{L^{-}L^{2}}\\
&=\frac{1}{\pi\alpha'}\frac{1}{I} ,
\end{split}
\end{equation}
where
\begin{equation}
\label{eq57}
\ I=\int^{z_{h}}_{0}dz\frac{1}{q(z)}\frac{z^{2}}{R^{2}\sqrt{f(z)[h(z)-f(z)]}}.
\end{equation}

The jet quenching parameter of an isotropic SYM result at zero magnetic field \cite{Liu:2006ug} has been given by
\begin{equation}
\label{eq58}
\ \hat{q}_{SYM}=\frac{\sqrt{\lambda} T^{3} \pi^{\frac{3}{2}} \Gamma(\frac{3}{4})}{\Gamma(\frac{5}{4})},
\end{equation}
where $ \sqrt{\lambda}=\sqrt{g^{2}_{YM}N_{c}}=\frac{R^{2}}{\alpha'}$.

By considering the jet moving along the $x_{1}$ direction and the momentum broadening occurring along the $x_{2}$ direction, we will study the second situation. One chooses the following static gauge:
\begin{equation}
\label{eq59}
\ x^{-}=\tau,\ x_{2}=\sigma,\  x^{+}=x_{3}=const.
\end{equation}

The jet quenching parameter is calculated as
\begin{equation}
\label{eq60}
\begin{split}
\hat{q}_{\perp(\perp)}&=8\sqrt{2}\frac{S_{I}'}{L^{-}L^{2}}\\
&=\frac{1}{\pi\alpha'}\frac{1}{I_{\perp(\perp)}} ,
\end{split}
\end{equation}
where
\begin{equation}
\label{eq61}
\ I_{\perp(\perp)}=\int^{z_{h}}_{0}dz\frac{1}{h(z)}\frac{z^{2}}{R^{2}\sqrt{f(z)[h(z)-f(z)]}}.
\end{equation}

\subsection{Moving parallel to the magnetic field}
By considering the jet moving along the $x_{3}$ direction and the momentum broadening occurring along the $x_{1}$ direction, we will study the jet quenching parameter of the parallel case. One chooses the following static gauge:
\begin{equation}
\label{eq62}
\ x^{-}=\tau,\ x_{1}=\sigma,\  x^{+}=x_{2}=const.
\end{equation}

The jet quenching parameter of the parallel case is
\begin{equation}
\begin{split}
\label{eq63}
\hat{q}_{\parallel(\perp)}&=8\sqrt{2}\frac{S_{I}''}{L^{-}L^{2}}\\
&=\frac{1}{\pi\alpha'}\frac{1}{I_{\parallel(\perp)}} ,
\end{split}
\end{equation}
where
\begin{equation}
\label{eq64}
\ I_{\parallel(\perp)}=\int^{z_{h}}_{0}dz\frac{1}{h(z)}\frac{z^{2}}{R^{2}\sqrt{f(z)[q(z)-f(z)]}}.
\end{equation}

\subsection{The calculated results of jet quenching parameter in a magnetized background}
\begin{figure}
    \centering
    \includegraphics[width=8.5cm]{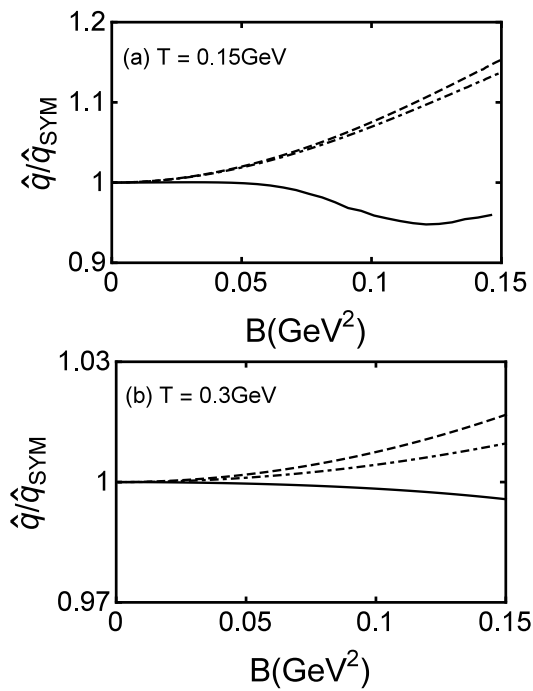}
    \caption{\label{fig5} Anisotropic effect on the magnetic field dependence of the jet quenching parameter normalized by the isotropic SYM result at zero magnetic field (a) for a low temperature ($T = 0.15 \textrm{GeV}$) and (b) for a higher one ($T = 0.3 \textrm{GeV}$). The dashed line indicates $\hat{q}_{\perp(\perp)}/\hat{q}_\textrm{SYM}$, the dash-dotted line indicates $\hat{q}_{\perp(\parallel)}/\hat{q}_\textrm{SYM}$, and the solid line denotes $\hat{q}_{\parallel(\perp)}/\hat{q}_\textrm{SYM}$.}
\end{figure}

In order to study the effects on the magnetic field dependence of the jet quenching parameter, we study the dependence of $\hat{q}/\hat{q}_\textrm{SYM}$ on a magnetic field for various temperatures in Fig.~\ref{fig5}. It is found that the jet quenching parameter when the probe is moving perpendicular to magnetic field direction is larger than when it is moving parallel to the magnetic field direction. The result suggests that the magnetic field inclines to suppress more jets in the transverse direction than in the parallel direction.

\begin{figure}
    \centering
    \includegraphics[width=8.5cm]{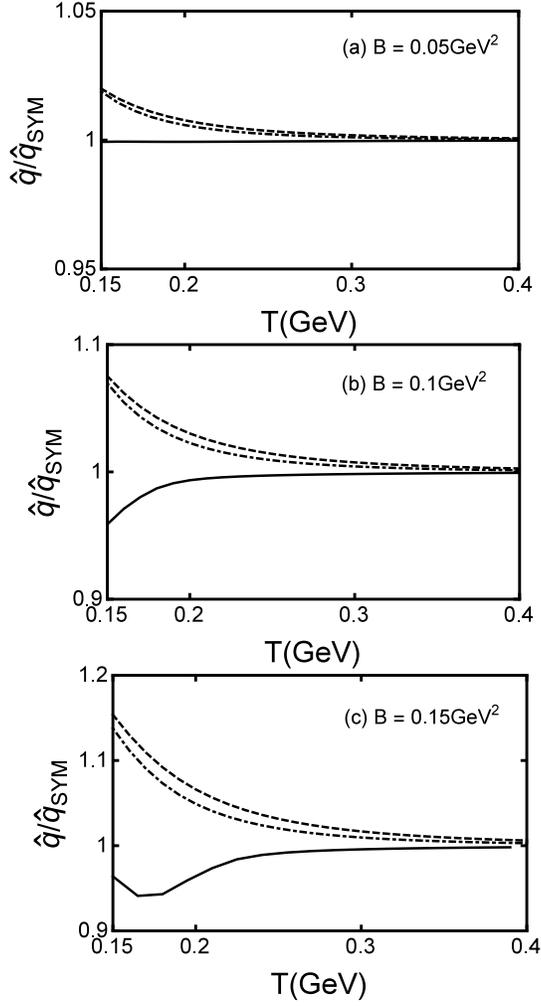}
    \caption{\label{fig6} Anisotropic effect on temperature dependence of the jet quenching parameter in the magnetized background normalized by the isotropic SYM result at zero magnetic field (a) for $B = 0.05 \textrm{GeV}^2$, (b)  for $B = 0.1 \textrm{GeV}^{2}$, and (c) for $B = 0.15 \textrm{GeV}^{2}$. The dashed line indicates $\hat{q}_{\perp(\perp)}/\hat{q}_\textrm{SYM}$, the dash-dotted line indicates $\hat{q}_{\perp(\parallel)}/\hat{q}_\textrm{SYM}$, and the solid line denotes $\hat{q}_{\parallel(\perp)}/\hat{q}_\textrm{SYM}$.}
\end{figure}

The anisotropic effect on the temperature dependence of the jet quenching parameter $\hat{q}/\hat{q}_\textrm{SYM}$ in the magnetized  background is given in Fig.~\ref{fig6}. One can see that the jet quenching parameter in the magnetized  background is obviously temperature dependent when $T \leq 0.3 \textrm{GeV}$, which shows that the magnetic field can change the isotropic phase of jet quenching at low temperature. But at relatively high temperature ($T > 0.3$), $\hat{q}/\hat{q}_\textrm{SYM}$ always tends to be 1, which means that the magnetic field hardly changes the isotropic phase of jet quenching. The authors of Ref. \cite{Zhang:2018mqt} investigated the dependence of a jet quenching parameter on $B/T^{2}$ at large magnetic field but did not study the dependencies of a jet quenching parameter on a magnetic field and temperature, separately. The dependencies of a jet quenching parameter on a magnetic field and temperature have been studied in the article.

\section{The energy loss of light quarks in a magnetized background}\label{sec:06}
In this section, we will pay attention to the energy loss of the light quarks in a magnetized background. A light quark-antiquark pair by an initially pointlike open string created with endpoints that propagate away from each other and fall from the boundary toward the horizon has been discussed by using the falling string\cite{Chesler:2008wd}. The endpoints must travel transversely to the string. It means no endpoint momentum or say there is no part of the string has momentum upward toward the boundary during the process of moving. But in\cite{Ficnar:2013qxa,Ficnar:2013wba}, there is a development for the endpoint’s trajectory that starts close to the horizon rather than close to the boundary in the  previous treatments of the falling string model.

The authors of Ref.\cite{Ficnar:2013wba} considered the endpoints representing the energetic quarks which were created though a hard scattering event and the string representing the color field generated by the quarks. The string is to start with packs of energy into the endpoints which were located close to the horizon. It is termed finite-endpoint-momentum shooting string(“the shooting string” for short): adding finite momentum at the endpoints and the trajectories of the endpoints follow the null geodesics while the rest of the string sags behind the endpoints. One can consider the endpoints are close to the horizon initially and “shot” towards the boundary. As it first reaches to a minimal value $z_{\ast}$ of the radial coordinate, it falls back into the horizon. There is a shooting limit that the energy loss can be approximated by using the null geodesic with $L$ = 1.  During the process of endpoints rising, the energy and momentum are gradually bled off. The string performs the process of a “snap-back” or “rising-and-falling”. It may offer a more natural description of the energetic light quarks passing through the QGP medium. Meanwhile the finite momentum endpoints can travel further than in the falling string model.

A general expression of the instantaneous energy loss of light quarks proposed in\cite{Ficnar:2013qxa}is taken as
\begin{equation}
\label{eq65}
\ \frac{dE}{dx}=-\frac{|L|}{2\pi\alpha'}G_{xx}(z),
\end{equation}
where the constant $L$ represents the null geodesics  that the endpoint follows. In order to make  Eq.(\ref{eq65}) into an usable expression, one needs to relate $L$ to some observable quantity and express $z$ in terms of $x$. From the Eq.(\ref{eq65}) one can see that if the endpoints starts close to the boundary($z$ is very small), the energy loss will be large. It means the jets will be quenched in a short time and will not be observable. So it needs to start close to the horizon in order to be observable.

\subsection{Moving perpendicular to the magnetic field}
By assuming the light quark moves in the $x_{1}$ direction and magnetic field along the $x_{3}$ direction and using metric $(2)$, we study the energy loss of light quarks in a magnetized background through the EM system with Eq.(\ref{eq2}). The energy and momentum of a probe are conserved as
\begin{equation}
\label{eq66}
\ E=-\frac{1}{\eta} \frac{R^2}{z^2}f(z),
\end{equation}
and
\begin{equation}
\label{eq67}
\ p_{x_{1}}=\frac{1}{\eta} \frac{R^2}{z^2}h(z) \frac{dx}{dt}.
\end{equation}
where $\eta(t)$ is an auxiliary field. The null geodesics $L$ read as
\begin{equation}
\label{eq68}
\ L=\frac{E}{p_{x_{1}}}=- \frac{f(z)dt}{h(z)dx}.
\end{equation}

The finite momentum endpoints will move along null geodesics $ds^{2}=0$, and one gets
\begin{equation}
\label{eq69}
\ (\frac{dx}{dz})^{2}=\frac{1}{h(z)[h(z)L^{2}-f(z)]}.
\end{equation}

If the endpoints follow the null geodesics, then the denominator of Eq.(\ref{eq69}) vanishes at $z=z_{\ast}$. One can relate it to $L$ as
\begin{equation}
\label{eq70}
\ L=-\sqrt{\frac{f(z_{\ast})}{h(z_{\ast})}}.
\end{equation}

Then one can get the energy loss as
\begin{equation}
\label{eq71}
\ \frac{dE}{dx}=-\frac{\sqrt{\lambda}}{2\pi}\frac{h(z)}{z^{2}}\sqrt{\frac{f(z_{\ast})}{h(z_{\ast})}},
\end{equation}
and
\begin{equation}
\label{eq72}
\ \frac{dx}{dz}=\sqrt{\frac{h(z_{\ast})}{h^{2}(z)f(z_{\ast})-f(z)h(z)h(z_{\ast})}}.
\end{equation}

As one assumes that the endpoint is at $x = 0$ and $z = z_{h}$ as the initial moment, then the endpoint is shot towards the boundary and reaches to a minimal value $z_{\ast}$ of the radial coordinate. Note that $z_{\ast}$ is very small. In the shooting limit, $z_{\ast}\rightarrow 0$, and large $x$ corresponds to the fact that the endpoint is closer to the boundary.

In order to calculate the energy loss $dE/dx$, we use the numerical simulation method to solve Eq.(\ref{eq72}) in the next section. In order to compare with the conformal case, we compute the ratio of energy loss with the magnetic field to the conformal field. The conformal case \cite{Ficnar:2013qxa} is

\begin{equation}
\label{eq73}
\ (\frac{dE}{dx})_{SYM}=-\frac{\pi\sqrt{\lambda}}{2}T^{2}(1+\pi T x)^{2}.
\end{equation}
where $ \sqrt{\lambda}=\sqrt{g^{2}_{YM}N_{c}}=\frac{R^{2}}{\alpha'}$.
\subsection{Moving parallel to the magnetic field}

We assume the light quark moves in the $x_{3}$ direction which is parallel to the direction of the magnetic field. The energy and momentum of a probe are conserved in metric (\ref{eq2})

\begin{equation}
\label{eq74}
\ E=-\frac{1}{\eta} \frac{R^2}{z^2}f(z),
\end{equation}
and
\begin{equation}
\label{eq75}
\ p_{x_{3}}=\frac{1}{\eta} \frac{R^2}{z^2}q(z) \frac{dx}{dt},
\end{equation}
where $\eta(t)$ is an auxiliary field.

By using the same procedure as above, one can get
\begin{equation}
\label{eq76}
\ \frac{dE}{dx}=-\frac{\sqrt{\lambda}}{2\pi}\frac{q(z)}{z^{2}}\sqrt{\frac{f(z_{\ast})}{q(z_{\ast})}},
\end{equation}
and
\begin{equation}
\label{eq77}
\ \frac{dx}{dz}=\sqrt{\frac{q(z_{\ast})}{q^{2}(z)f(z_{\ast})-f(z)q(z)q(z_{\ast})}}.
\end{equation}

\subsection{The calculated results of the shooting string}
\begin{figure}
    \centering
    \includegraphics[width=8.5cm]{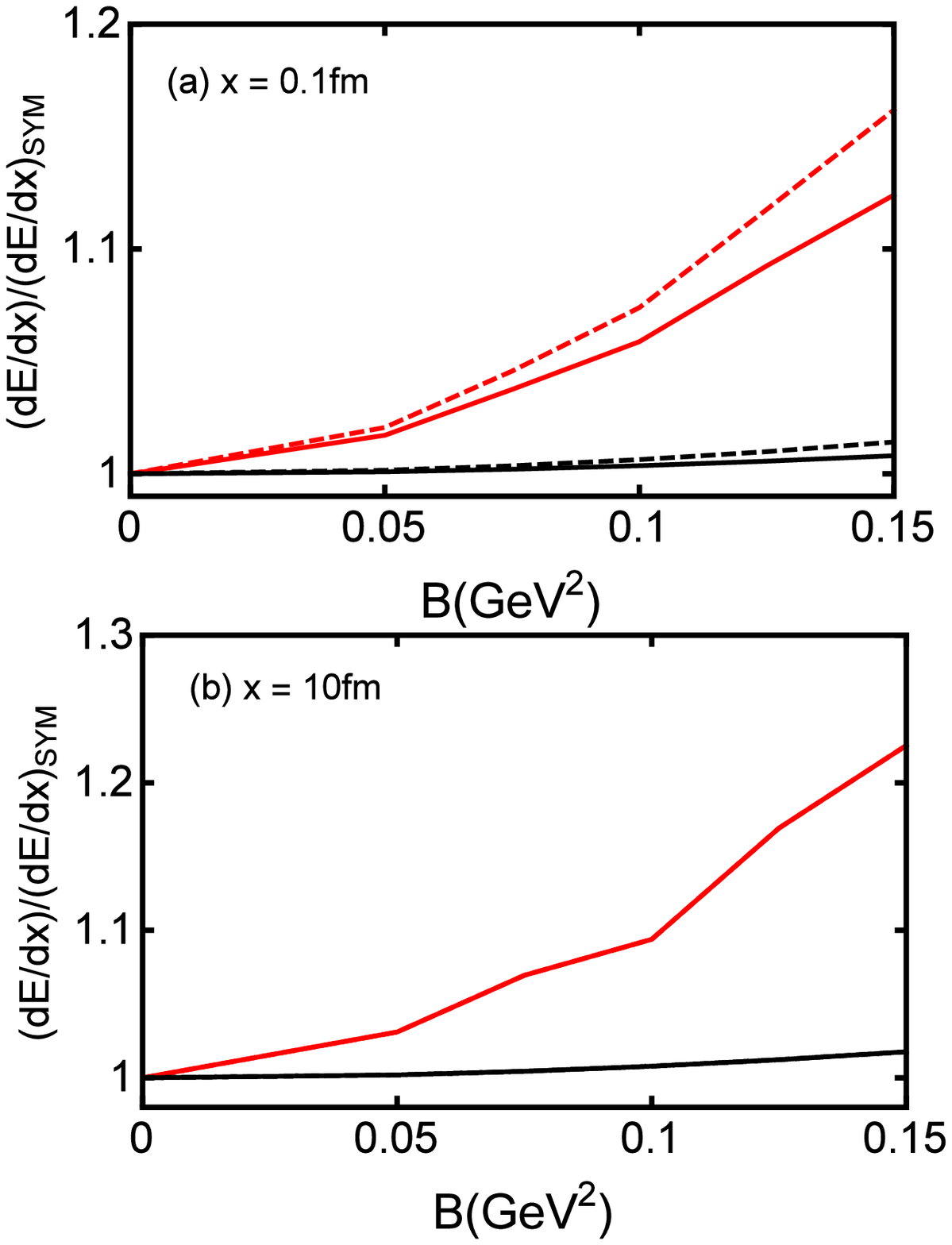}
    \caption{\label{fig7} Anisotropic effect on the magnetic field dependence of the energy loss of light quarks normalized by the isotropic SYM result at zero magnetic field (a) for a small distance $x$ ($x = 0.1fm$) and (b) for a bigger one($x = 10fm$). The solid line indicates the light quark moving parallel to the magnetic field direction, and the dashed line indicates the light quark moving perpendicular to the magnetic field direction.The red line denotes a low temperature($T = 0.15 \textrm{GeV}$) and the black line denotes a higher one($T = 0.3 \textrm{GeV}$).}
\end{figure}

In a search for the holographic effect on a magnetic field dependence of the energy loss of light quarks, the dependence of $(dE/dx)/(dE/dx)_\textrm{SYM}$ on the magnetic field at some fixed temperatures and $x$ are provided in Fig.~\ref{fig7}. It is found that the energy loss of light quarks increases with the magnetic field continuously, and it is larger than that of $N=4$ SYM theory especially at low temperature. It suggests that the magnetized background enhances the energy loss of the light quarks. One also finds that in early times (small $x$), the effects on the energy loss of light quarks when moving perpendicular to the magnetic field direction is more obvious than moving parallel to the magnetic field direction, which implies that the magnetic field tends to suppress more quarks and jets when moving in the transverse direction than in the parallel direction in the early times. But in later times(large $x$), the effects on energy loss of light quarks of moving perpendicular to the magnetic field direction is nearly the same as that of moving parallel to the magnetic field direction, which suggests that the effect on the magnetic field of the energy loss in the transverse direction is almost the same as that moving in the parallel direction in the later times.

\begin{figure}
    \centering
    \includegraphics[width=8.5cm]{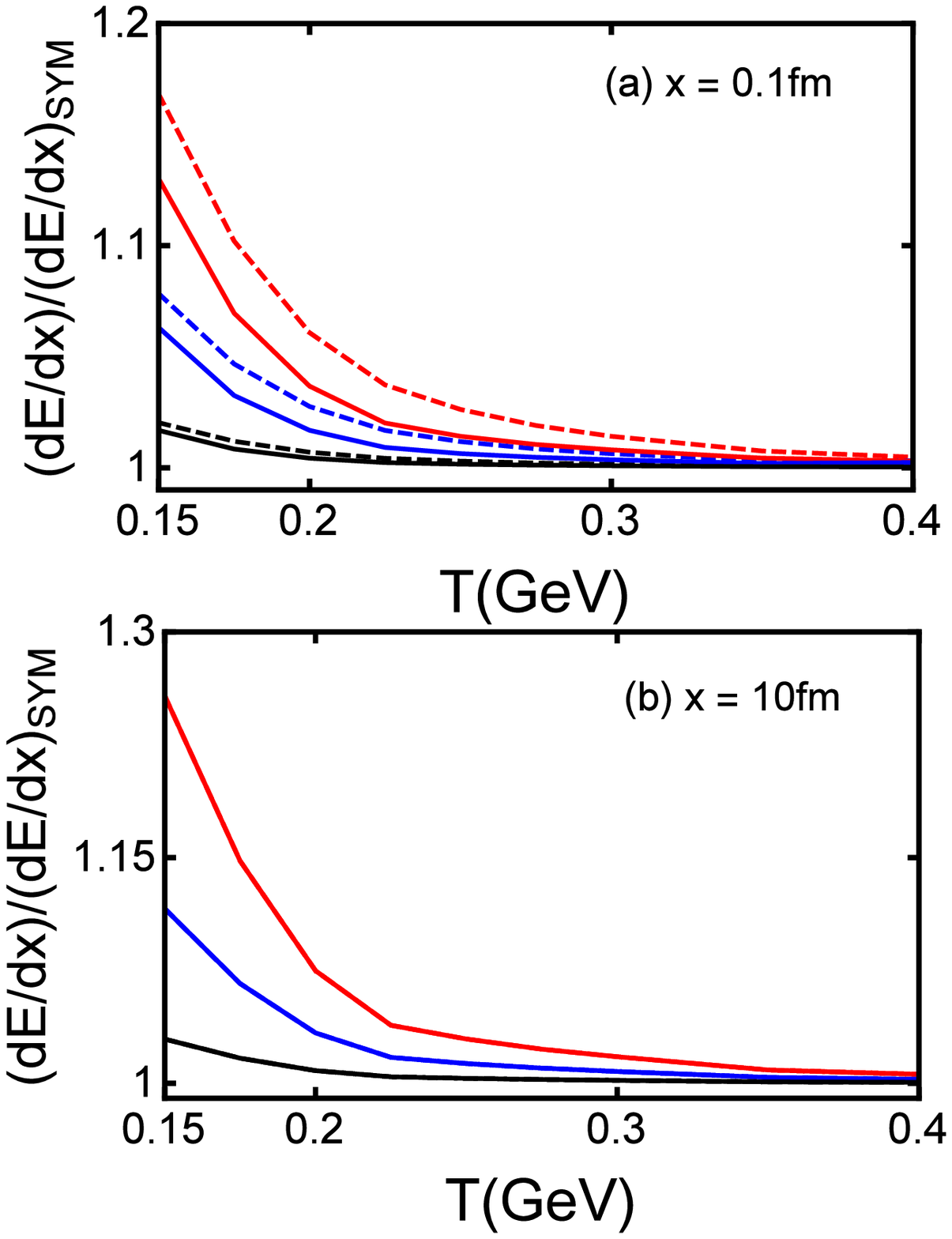}
    \caption{\label{fig8}   Anisotropic effect on temperature dependence of the energy loss of light quarks in the magnetized background normalized by the isotropic SYM result at zero magnetic field. (a) for a small distance $x$($x = 0.1fm$) and (b) for a bigger one($x = 10fm$). The solid line indicates the light quark moving parallel to the magnetic field direction, and the dashed line indicates the light quark moving perpendicular to the magnetic field direction. The red line denotes $B = 0.15 \textrm{GeV}^{2}$ , the blue line denotes $B = 0.1 \textrm{GeV}^{2}$ and the black line denotes $B = 0.05 \textrm{GeV}^{2}$.}
\end{figure}

Figure.~\ref{fig8} shows the effect on temperature dependence of the energy loss of light quarks $(dE/dx)/(dE/dx)_\textrm{SYM}$ for various magnetic field and distance $x$. The ratio decreases rapidly with the increase of temperature at $T \leq 0.3$, which expresses a strong temperature dependence at low temperature. But at relatively high temperature ($T > 0.3$), the ratio always tends to be 1, which means the magnetic field hardly changes the isotropic phase at high temperature.

\section{Conclusion and discussion}\label{sec:07}
Some RHIC and LHC data indicate the production of jet quenching and a strong magnetic field. As is well known noncentral heavy ion collisions at the RHIC and the LHC can produce a very strong magnetic field. Besides the phenomenological significance in theoretical sides of strong interaction, strong magnetic fields in relativistic heavy-ion collisions also provide some deep investigations of the dynamics of QCD, of which the vacuum structure is of numerous interest. The magnetic field is introduced into the background geometry by solving the Einstein-Maxwell system in this paper. After embedding the magnetized background geometry into the modified soft-wall model, we study the magnetic field dependent behaviors of jet quenching and energy loss numerically.

In summary, the holographic effects on the magnetic field dependence of the drag force, diffusion coefficient, jet quenching parameter of heavy quarks, and the energy loss of the light quarks were systematically studied in the paper. It was found that the perpendicular case is larger than that of the parallel case, which implies that the magnetic field tends to suppress more quarks and jets in the transverse direction than in the parallel direction. One also finds that the magnetic field will enhance the energy loss of the light quarks. The diffusion coefficient decreases with the magnetic field in the transverse direction, but in the high temperature and velocity case the diffusion coefficient increases with the magnetic field when moving parallel to the magnetic field direction, which indicates that the quark may diffusion farther when moving along the parallel to the magnetic field direction.

The dependencies of energy loss of heavy and light quarks on temperature for various magnetic fields were studied in this article. We found some important phenomena with the increase of temperature. When the temperature rises to $0.3 \textrm{GeV}$, the ratio approaches to 1, which suggests that the magnetic field hardly changes the isotropic phase in high temperature.

\section*{ACKNOWLEDGMENTS}
This work was supported by National Natural Science Foundation of China (Grants No. 11875178, No. 11475068, and No. 11747115) and the CCNU-QLPL Innovation Fund (Grant No. QLPL2016P01).

\section*{References}

\bibliography{ref}
\end{document}